\documentstyle[12pt,epsfig]{article}
\newcommand{\be}{\begin{equation}}
\newcommand{\ee}{\end{equation}}
\newcommand{\bea}{\begin{eqnarray}}
\newcommand{\eea}{\end{eqnarray}}

\begin{document}
\title{Probing hadronic formation times with antiprotons in $p+A$ reactions at AGS energies
     \thanks{Supported by the European Graduate School 'Complex Systems of Hadrons and Nuclei',
      Copenhagen-Giessen} }
\author{ W. Cassing$^1$, E. L. Bratkovskaya$^2$, O. Hansen$^3$\\
\\
    $^1$Institut f\"{u}r Theoretische Physik, Universit\"{a}t Giessen \\
    D-35392 Giessen, Germany\\
    $^2$Institut f\"{u}r Theoretische Physik, Universit\"{a}t Frankfurt
    \\
    D-60054 Frankfurt, Germany\\
    $^3$Niels Bohr Institute, University of Copenhagen \\  2100 Copenhagen, Denmark}
\maketitle
\begin{abstract}
The production of antiprotons in $p+A$ reactions is calculated in
a microscopic transport approach employing hadronic and string
degrees of freedom (HSD). It is found that the abundancies of
antiprotons as observed by the E910 Collaboration in $p+A$
reactions at 12.3 GeV/c as well as 17.5 GeV/c  can approximately
be described on the basis of primary proton-nucleon and secondary
meson-baryon production channels for all targets. The transport
calculations demonstrate that the antiproton rapidity
distributions for heavy targets are sensitive to the $\bar{p}$ (or
hadron) formation time in the nuclear medium. Within our analysis
the data from the E910 Collaboration  are reasonably described
with a formation time of $ 0.4-0.8$ fm/c in the hadron rest frame.
\end{abstract}

\vspace{2cm} \noindent PACS: 24.10.-i; 24.10.Cn; 24.10.Jv;


\newpage
\section{Introduction}
Since the first observation of antiproton production in
proton-nucleus \cite{chamberlain,elioff,dorfan} and
nucleus-nucleus collisions \cite{JINR,BEVALAC1,BEVALAC2}
the production mechanism has been quite a matter of debate. Especially
in nucleus-nucleus collisions at subthreshold energies
traditional cascade calculations, that employ free $NN$ production
and $\bar{p}N$ annihilation cross sections, essentially fail in
describing the high cross sections seen from 1.5 - 2.1 A$\cdot$GeV
\cite{Cass92,Faess1}. Thus multiparticle nucleon
interactions \cite{Danielewicz90,Weise} have been suggested as a
possible solution to this problem. On the other hand it has been
pointed out  that the quasi-particle properties of the nucleons
and antinucleons  might be important for the $\bar{p}$ production
process which become more significant with increasing nuclear
density \cite{Schaffner91,Koch,Teis94,Ko93,RQMD,sibirtsev}.
Relativistic transport calculations have indicated that
all the low energy data from proton-nucleus and nucleus-nucleus
collisions are compatible with attractive $\bar{p}$ self energies
(at normal nuclear matter density $\rho_0$) in the order of -100
to -150 MeV \cite{sibirtsev,Cass99,Ko96}. However, it had been
stressed at that time that the high antiproton yield might also be
attributed to mesonic production channels  \cite{ko1}
since $p \bar{p}$ annihilation leads to multi-pion final states
with an average abundancy of 5 pions \cite{Dover}.

With new data coming up on antibaryon production from
proton-nucleus and nucleus-nucleus collisions at the AGS
\cite{AGSall,E877,AGSnew,E864n,E910} and SPS
\cite{NA49,Na49b,NAxx,Andersen} the $\bar{p}, \bar{\Lambda}$
enhancement factors seen experimentally were no longer that
dramatic as at SIS energies, however, traditional cascade
calculations employing free production and annihilation cross
sections again were not able to reproduce the measured abundancies
and spectra \cite{AGS1,AGS2,Kahana,Bleicher} especially for $\Xi,
\bar{\Xi}$ and $\Omega, \bar{\Omega}$. Here additional collective
mechanisms in the entrance channel have been suggested such as
color rope formation \cite{Sorge} or hot plasma droplet formation
\cite{Werner}. In another language this has been addressed also as
string fusion \cite{Carlos,Carlos2}, a precursor phenomenon for
the formation of a quark-gluon plasma (QGP).

The intimate connection of antibaryon abundancies with the
possible observation of a new state of the strongly interacting
hadronic matter, i.e. the quark-gluon plasma, has been often
discussed since the early suggestion in Ref. \cite{Rafelski} that
especially the enhanced yield of strange antibaryons --
approximately in chemical equilibrium with the other hadronic
states -- should be a reliable indicator for a new state of
matter. In fact, the data on strange baryon and antibaryon
production from the NA49 and WA97 Collaborations show an
approximate chemical equilibrium \cite{BM,becca,Red01} with an
enhancement of the $\Omega^-, \Omega^+$ yield in central $Pb + Pb$
collisions (per participant) relative to $p Be$ collisions at the
same invariant energy per nucleon by a factor $\sim $ 15. As
pointed out in Ref. \cite{Redlich2} the data on multi-strange
antibaryons at the SPS seem compatible with a canonical ensemble
in chemical equilibrium.

In Ref. \cite{Rapp} Rapp and Shuryak have proposed multi-meson
production channels for $p\bar{p}$ pairs  to describe the
antiproton abundancies in central $Pb+Pb$ collisions at the SPS.
Later on, Greiner and Leupold \cite{Carsten} have applied the same
concept for the $\bar{\Lambda}$ production by a couple of mesons
including a $K^+$ or $K^0$ (for the $\bar{s}$ quark). In fact,
their estimates have been supported by the microscopic
multi-particle calculations in Ref. \cite{Cass02} employing
detailed balance relations for 2 hadron $\leftrightarrow$  n
hadron transitions and vice versa. Thus, in case of relativistic
nucleus-nucleus collisions at AGS and SPS energies the low
effective antibaryon absorption seen experimentally can naturally
be explained by the backward reaction channels, i.e. multi-meson
fusion  channels.

The latter mechanism, however, does not apply to proton-nucleus
reactions since there is no longer an approximately thermal 'meson
bath' feeding the baryon-antibaryon production channel as pointed
out in Ref. \cite{Cass02}. Nevertheless, the $\bar{p}$ yield
measured by the E910 Collaboration in $p+A$ reactions at 12.3
GeV/c and 17.5 GeV/c was found to be surprisingly high
\cite{E910}. An analysis within the Glauber model showed that
either the annihilation cross section $\sigma_{ann}$ should be
small compared to the cross section in free space or the $\bar{p}$
formation time $\tau_F$ should be a couple of fm/c when employing
the free annihilation cross section \cite{E910}. The authors of
the latter work conclude that the 'naive' first-collision model
with subsequent annihilation of antiprotons is incompatible with
their data supporting the earlier finding from Ref. \cite{Gonin},
where previous data from $p+A$ and $A+A$ reactions had been
analysed within a similar model.

In this work we will address $\bar{p}$ production in
proton-nucleus collisions at 12.3 GeV/c and 17.5 GeV/c within the
covariant HSD\footnote{Hadron-String-Dynamics} transport approach
\cite{Ehehalt} which is based on the concept of string formation
and decay at invariant collision energies $\sqrt{s} \geq$ 2.6 GeV
for baryon-baryon collisions and $\sqrt{s} \geq$ 2.3 GeV for
meson-baryon collisions, while the low-energy reactions are
described by known vacuum cross sections or resonance formation
and decay cross sections. A single sensitive parameter in this
approach is the string duration or hadron formation time $\tau_F
\approx$ 0.7-0.8 fm/c which has been fitted to the proton rapidity
distribution in nucleus-nucleus collisions at SPS energies
\cite{Ehehalt} and kept fixed furtheron. For more details we refer
the reader to the review \cite{Cass99}.

\section{Qualitative considerations}
The physical picture within this transport approach for $\bar{p}$ production in $p+A$
reactions is as follows: The impinging proton might create a
$N\bar{N}$ pair in a first collision at position $x_1$ (cf. Fig.
1) and the prehadronic quark-antiquark configuration travels essentially in
forward direction by a time $\gamma \tau_F$ -- where $\gamma$
denotes the Lorentz dilatation factor -- until the hadronic
$\bar{p}$ emerges at position $x_2$. The hadronic state then
propagates further in the same direction and reacts with the
remaining nucleons elastically or inelastically. In the latter
phase the dominant channel will be annihilation with a cross
section
\begin{equation}
\label{appann} \sigma_{ann} (\sqrt{s}) = \frac{50 [mb]}{v_{rel}} ,
\end{equation}
where the relativistic relative velocity is given by
\cite{Byckling}
\be
\label{vrel} v_{rel}= \frac{\sqrt{(s-M_1^2-M_2^2)^2-4 M_1^2 M_2^2}
}{2 E_1 E_2}, \ee which is evaluated in the lab. frame. As shown
in Fig. 3 of Ref. \cite{Cass02} the expression (\ref{appann}) well
describes the experimental data \cite{LB,Caso} on $\bar{p}$
annihilation for relative momenta up to 2--3 GeV/c.

In rapidity space the produced antibaryons will be localized
around midrapidity $y_m$; fast antiprotons hadronize later than
slow $\bar{p}$'s due to the larger $\gamma$ factor and in view of
(\ref{appann}) also experience a smaller annihilation cross
section. Thus antiprotons from first-chance collisions in heavy
nuclei should result in rapidity distributions that are highly
asymmetric with respect to midrapidity and a larger suppression
should be observed for $y \leq y_m$ than for $y \geq y_m$, if the
target is located at $y=$ 0. In fact, a simple look at the
experimental $\bar{p}$ rapidity distribution from Fig. 2 of Ref.
\cite{E910} does not support the first-collision picture.

On the other hand, in Ref. \cite{sibirtsev} it was found that a
dominant fraction of antiprotons emerge from secondary
pion-nucleon reactions in $p+A$ reactions at lower energies. Such
secondary reaction channels might also be a reason for the
observations of Ref. \cite{E910} since the $\bar{p}$ production
cross section in $\pi N$ reactions close to threshold is large
compared to the production cross section in $pp$ collisions (cf.
Fig. 6 of Ref. \cite{sibirtsev}). The dymical effects of such
secondary production channels are easy to map out within the
string/hadron dynamical picture: In a first $pN$ collision a
'meson' ($q \bar{q}$ pair) $M$ is produced at position $x_1$ which
hadronizes after time $\gamma_M \tau_F$ at position $x_2$. Then it
propagates on average for a distance $\lambda_M \sim (\sigma_{tot}
\rho_0)^{-1} \approx$ 2 fm (in the lab. frame of reference) before
reacting with another nucleon. In this reaction (at position
$x_3$) it may also produce a pre-$N\bar{N}$ pair which hadronizes
after time $\gamma \tau_F$ in position $x_4$. Then the antinucleon
propagates in the same direction and may be rescattered,
annihilated or emerge to free space. The antibaryons from
secondary -- or even ternary -- processes have a much shorter way
to escape the nucleus without being annihilated than those from
first-chance collisions. Furthermore, their initial rapidity
distribution will be shifted closer to target rapidity because the
primary mesons are produced symmetrically around $y_m$  and only
the energetic mesons will give a sizeable contribution to the
$\bar{p}$ production provided that they hadronize in the medium.

 As mentioned above we
here adopt the HSD transport approach that has e.g. been tested in
Ref. \cite{Geiss} for pion production and proton stopping in $p+A$
reactions at AGS energies using $\tau_F$ = 0.8 fm/c. Since the
description of the related data is very satisfactory (cf. Fig. 8
in Ref. \cite{Geiss}) we continue with aspects of meson formation
and secondary interactions, that are relevant for antiproton
production and propagation in $p+A$ reactions.

In order to achieve a quantitative idea about the time scales for
meson production and propagation we show in Fig. 2 for $p+Au$ at
12.3 GeV/c the time dependence of the 'formed' meson number (solid
line) in comparison to the number of string-like $q\bar{q}$
excitations (dashed line) that convert to mesons on a time scale
of $\tau_F$ in their individual rest frame. The reference frame
for the transport calculation is the nucleon-nucleon cms. (at
$y_{lab} \approx 1.63$) with a Lorentz $\gamma$-factor of $\approx
2.67$ for the projectile proton and target nucleus, respectively,
such that the dimensions of the target in space - as seen by the
hadrons - are squeezed by the factor $\gamma$. The sharp increase
of the dashed line at t= 0 denotes the beginning of the $p+Au$
reaction with the excitation of strings that may lead to the
formation of $N\bar{N}$ pairs, too. The latter also appear as
hadronic states on a similar time scale as the mesons --
approximately 2 fm/c later -- and may be annihilated or
rescattered on the residual target nucleons. We recall that only
the formed hadronic states are allowed to interact with the
nucleons of the target again and to produce secondary mesons or
baryon-antibaryon pairs. Thus antiprotons or mesons, that move
fast with respect to the target at rest, may only form in the
vacuum and not interact in the nucleus again. In fact, this is
approximately the case for a $Be$-target at this energy, while the
dominant fraction of hadrons are formed inside the $Au$-nucleus
and reinteract again (see below). For completeness we note, that
the increase of the meson number for $t \geq 7$ fm/c is
essentially due to the decays of mesonic resonances ($\rho,
\omega, \phi, K^*$) that increase the net meson number.

The distribution in the invariant collision energy $\sqrt{s}$ is
shown in Fig. 3 for $p+Au$ at 12.3 GeV/c for baryon-baryon ($BB$)
collisions (solid line) in comparison to the reactions of formed
mesons with nucleons ($mB$, dashed line). The high energy peak of
the $BB$-distribution (around $\sqrt{s} \approx$ 5 GeV) is due to
primary $pN$ collisions; it is smeared out substantially due to
Fermi motion. The low energy peak of the $BB$-distribution
reflects multiple scattering of baryons in the target that,
however, do not contribute to $p\bar{p}$ production  due to the
threshold of $\sim$ 3.75 GeV (solid arrow). The $\sqrt{s}$
distribution for interactions of formed mesons with baryons
(dashed line) extends to about 4.5 GeV, which is well above the
$\bar{p}p$ production threshold of $\sim$ 2.8 GeV (dashed arrow).
The distribution in the collision number above threshold is
substantially larger for $BB$-reactions than for meson-nucleon
collisions, however, the distributions have to be 'folded' with
the respective $p\bar{p}$ production cross sections. We recall
that in $\pi N$ reactions close to threshold this cross section is
large compared to the production cross section in $pp$ collisions
(cf. Fig. 6 of Ref. \cite{sibirtsev}) such that the $mB$ channels
might well compete with the $BB$ channels.

\section{Comparison to data}
It should be noted that for the present analysis the perturbative
production scheme of Ref. \cite{sibirtsev} has been employed due
to the low multiplicities for $\bar{p}$ production ($\sim 2\times
10^{-4}$ for $p+Au$ at 12.3 GeV/c). All cross sections are the
same as in the latter work; the only modification introduced is a
new parametrisation of the individual rapidity $(y)$ and
transverse momentum $(p_T)$ dependence for antiprotons in the
elementary $pN$ and $\pi N$ production channels, since the
kinematical regime is substantially different from the latter
study and the final states can no longer be described accurately
by 4- or 3-body phase-space, respectively. To this aim high
statistics event generation has been performed within the LUND
string model \cite{LUND} -- as implemented in the HSD approach --
for $\bar{p}$ production and the resulting two-dimensional spectra
have been fitted by a Gaussian in rapidity and exponential in
$p_T^2(y)$. As demonstrated in Refs. \cite{Cass99,Geiss} the LUND
model works rather well for this purpose in comparison to the
available data \cite{Caso}, however, uncertainties in the order of
50\% for the elementary cross sections cannot be ruled so far
since there are no data available for invariant energies close to
threshold.

With the input cross sections defined we can step on with a
quantification of the physics addressed above. In Fig. 4 we show
the $\bar{p}$ rapidity spectra for $p+Be$ at 12.3 GeV/c in
comparison to the experimental data from \cite{E910}. The dashed
line ($BB prim.$, upper part) gives the $\bar{p}$ spectrum from
$BB$ production channels prior to final state interactions (FSI).
The contribution from secondary $mB$ collisions is not visible on
a linear scale for the $Be$ target. When including $\bar{p}$
rescattering as well as annihilation the dash-dotted line ($BB
final$) is obtained for $BB$ production channels. It is clearly
seen that  $\bar{p}$ annihilation is more pronounced for low
rapidities -- with the target at $y=0$ -- than for $y \geq y_m
\approx $ 1.63. Thus for $p+Be$ the $pN$ production channel
clearly dominates and the comparison with the experimental
rapidity distribution (lower part of Fig. 4) essentially tests the
adequacy of the elementary differential $\bar{p}$ production cross
section in $pN$ collisions which appears to be underestimated
slightly (by about 30\%) in the present approach using the input
from Ref. \cite{sibirtsev} and assuming that the data point for
$y$ = 1.1 is solid. One might rescale the elementary cross section
accordingly, however, the focus of our present work is not on
fitting data. On the other hand, antiprotons from a feeddown of
$\bar{\Lambda}$ and $\bar{\Sigma}^0$ might also be contained in
the data while they are excluded in the calculation\footnote{In
Ref. \cite{E910} such a feeddown contribution was estimated to be
less than 5\%.}. Thus, provided that the reactions are described
properly in the HSD transport approach, the cross sections for all
targets should be low by about 30\%, too, in case the $\bar{p}$
yield is dominated by the $BB$ channels (see below).

The numerical results for $p+Cu$ and $p+Au$ at 12.3 GeV/c are
shown in Figs. 5 and 6 adopting the same assignment of the lines
as in Fig. 4. It is clearly seen that for the $Cu$-target the
contribution from $mB$ collisions -- as displayed by the short
dashed line ($mB prim.$) --
becomes visible and that the
annihilation of $\bar{p}$'s from the secondary interactions is
rather small as expected from the schematic Fig. 1. On the other hand, in case of the
$Au$-target the antiprotons from $NN$ channels are annihilated to
a large extent especially for $y \leq y_m$ and the surviving
$\bar{p}$'s in the experimental rapidity window $1.0 \leq y \leq
2.0$ \cite{E910} practically stem from the secondary production channels.
Nevertheless, the experimental data are underestimated on average
again by about 30\% as in case of the $Be$-target which might be due to
improper elementary cross sections and/or still missing production
channels in the transport approach.

The relative antiproton absorption according to Figs. 4--6
strongly depends on rapidity due to a larger formation time
$\tau_F$ and lower annihilation cross section with increasing
rapidity $y$. In case of $p+Be$ there is practically no $\bar{p}$
annihilation for $y \geq$ 2 since the fast antinucleons only form
in the vacuum. For the $p+Cu$ system this threshold is shifted to
$y \geq$ 2.6 while for the $p+Au$ system the annihilation becomes
negligible only for $y \geq$ 3.  The sensitivity to the formation
time $\tau_F$ is demonstrated in the lower part of the Figs. 5 and
6, where also the final antiproton rapidity spectra are displayed
for $\tau_F$ = 0.4 fm/c (dashed line) and 1.6 fm/c (dotted line),
respectively. In case of the light $Be$-target no sensitivity to
$\tau_F$ was found within the numerical statistics of the
transport calculations since there is practically no contribution
from the $mB$ channels. As can be extracted from Figs. 5 and 6 a
shorter formation time $\tau_F$ leads to an enhancement of the
secondary production channel by $mB$ collisions, whereas a larger
formation time ($ \sim$ 1.6 fm/c) essentially suppresses the
contribution from secondary meson induced reactions in line with
Fig. 1. In fact, the experimental spectra for $Cu$ and $Au$ appear
to be better described by $\tau_F$ = 0.4 fm/c for the elementary
cross sections employed.

The average $\bar{p}$ multiplicity in the rapidity interval $1
\leq y \leq 2$ at 12.3 GeV/c is compared in Fig. 7 as a function
of target mass $A$ with the data from \cite{E910} for different
$\tau_F$. As mentioned before, for the $Be$-target the
calculations for $\tau_F$ = 0.4, 0.8 and 1.6 fm/c give
approximately the same result, which in comparison to the
experimental data is down by about 30\% just scratching the lower
level of the experimental error bars. For the $Cu$-target both
calculations for $\tau_F$ = 0.8 fm/c and  $\tau_F$ = 1.6 fm/c fall
down too low. In case of the $Au$-target only the calculation for
$\tau_F$ = 1.6 fm/c is out of range in comparison to the data and
also incompatible with the shape of the $dN/dy$ spectra such that
a formation time in the order of 1.6 fm/c should be excluded.

We, furthermore, compare our calculations to the system $p+Au$ at
17.5 GeV/c where the E910 Collaboration has taken data, too
\cite{E910}. In this case (Fig. 8) the most important production
channel is due to primary $pN$ inelastic collisions, however, the
surviving antiprotons dominantly appear above midrapidity $y_{m}
\approx$ 1.81 due to the strong absorption. Within the rapidity
window $1 \leq y \leq 2$ of the data the contribution from
secondary $mB$ channels is almost compatible (thin solid line, $mB
final$, compared to the dot-dashed line, $BB final$). The sum of
the contributions is reasonably well in line with the experimental
data from \cite{E910} (lower part) for $\tau_F$ =0.8 fm/c (solid
line). A slightly better description is achieved with $\tau_F$ =
0.4 fm/c for $y \approx$ 1-1.5 in line with the data at 12.3
GeV/c, however, now the $\bar{p}$ rapidity distribution is
slightly overestimated for $y \approx$ 2. In view of the fact,
that the elementary antiproton production cross section is
underestimated by about 30\% in our calculations, this provides a
lower limit for the formation time $\tau_F$.

\section{Summary}
In this work we have analyzed antiproton production in $p+A$
reactions at AGS energies within the HSD transport approach to
examine the apparent low $\bar{p}$ annihilation seen
experimentally \cite{E910,Gonin}. Whereas the rather high
antibaryon yield in nucleus-nucleus collisions at AGS and SPS
energies can be attributed to multi-meson fusion channels
employing detailed balance \cite{Rapp,Carsten,Cass02} the
experimental data in proton-nucleus collisions at AGS energies
appear to be compatible with primary proton-nucleon and secondary
meson-baryon production channels as also found in Ref.
\cite{sibirtsev} for lower energies.

Detailed calculations employing different hadronic formation times
rule out long formation times of the order of 1.6 fm/c (in the
hadron rest frame) since such times strongly suppress the meson
reinteraction rate. The comparison with the data from the E910
Collaboration instead favors  formation times of 0.4-0.8 fm/c when
adopting the free antibaryon annihilation cross section. The
latter cross section might be somewhat reduced in the nuclear
medium, but it is not likely to be down by factors $\sim$ 5 as
quoted in the analysis of Ref. \cite{E910}. We have pointed out,
furthermore, that especially the high momentum/rapidity part of
the $\bar{p}$ spectrum, which dominantly arises from primary $pN$
collisions, shows a sensitivity to the $\bar{p}$ formation time
since 'fast' antiprotons may only form in free space and no longer
get annihilated in the medium. Unfortunately, this rapidity regime
is out of the E910 acceptance such that further experimental
studies will be necessary to provide a final answer to the issue
of antiproton formation times.

\vspace{0.5cm} The authors like to acknowledge valuable discussions with C. Greiner.

%----------------------------------------------------------------------

\newpage
\begin{figure}[h]
\centerline{\psfig{file=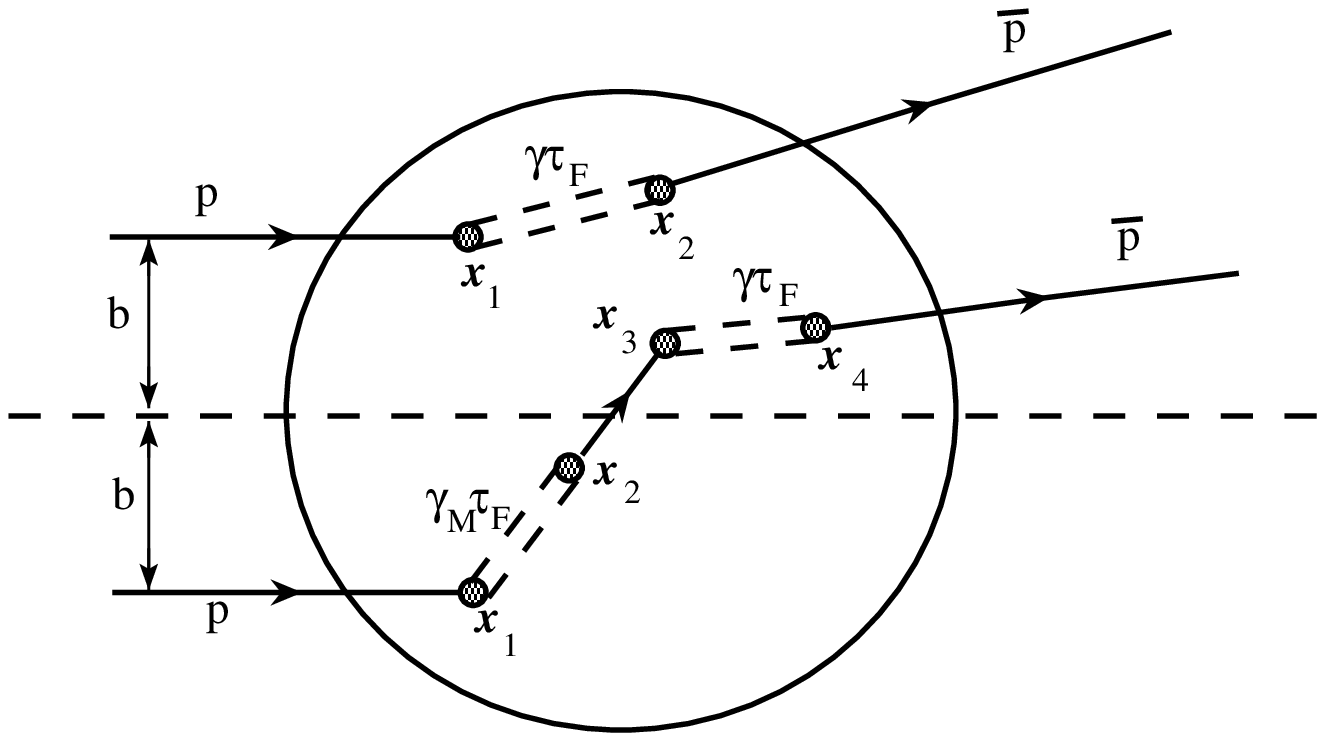,height=7cm}}
\caption{Illustration of primary $p\bar{p}$ production (in $x_1$),
formation (in $x_2$) and propagation in the nucleus (upper part).
The lower part shows the primary production of a pre-meson
($q\bar{q}$ pair) in $x_1$, its formation to a hadron in $x_2$,
the propagation of the hadron in the nucleus and rescattering with
another nucleon in $x_3$. In the secondary interaction also a
pre-$p\bar{p}$ can be created which hadronizes in $x_4$ and
further propagates in the nucleus or in the vacuum (see text).}
\label{bild1}
\end{figure}
\begin{figure}[h]
\centerline{\psfig{file=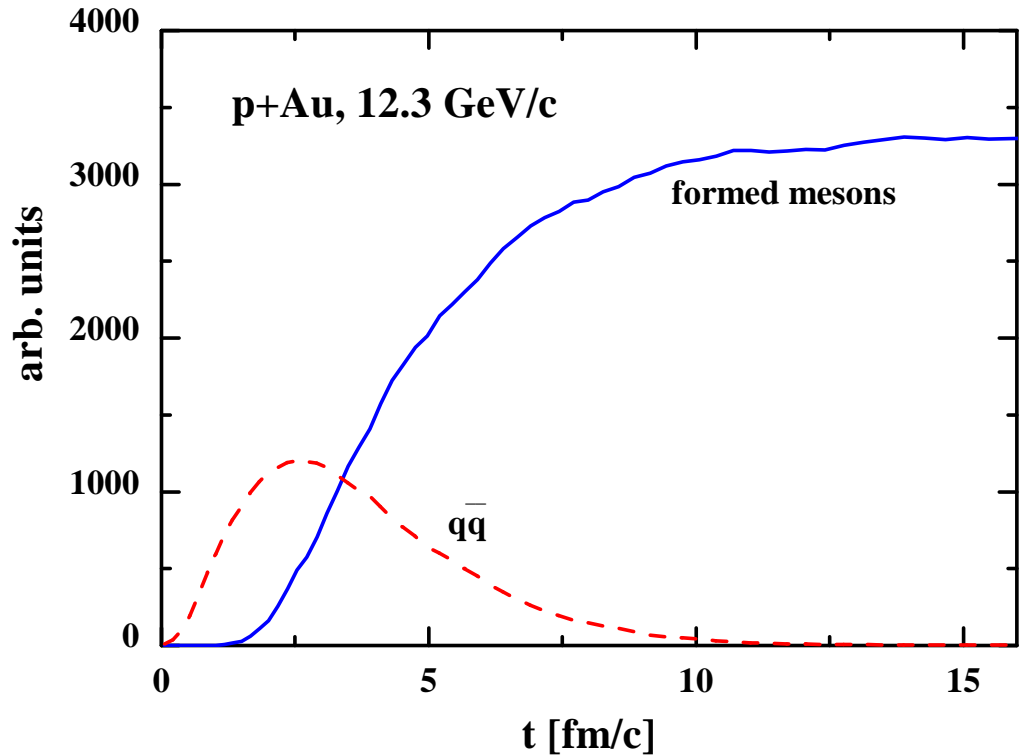,height=10cm}}
\caption{Time evolution of the string-like $q\bar{q}$ excitations (dashed line) and the
formed meson number in $p+Au$ reactions at 12.3 GeV/c within the HSD approach
employing the 'default' formation time $\tau_F$ = 0.8 fm/c.
The time is given in the $NN$ center-of-mass system.
}
\label{bild2}
\end{figure}
\begin{figure}[h]
\centerline{\psfig{file=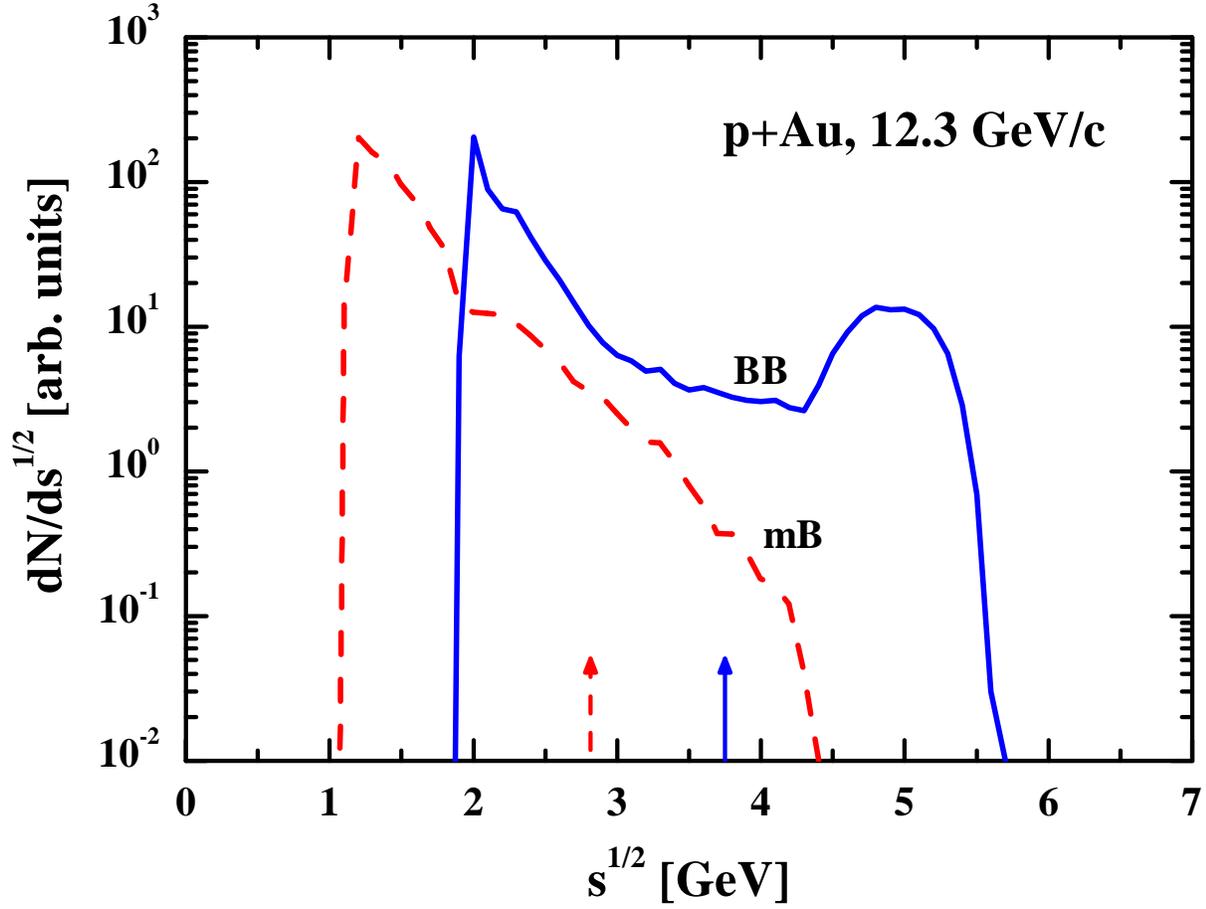,height=12cm}}
\caption{The differential number of baryon-baryon collisions (solid line) and meson-baryon
collisions (dashed line) in $p+Au$ reactions at 12.3 GeV/c within the HSD approach
employing the 'default' formation time $\tau_F$ = 0.8 fm/c.
The solid arrow denotes the threshold for $p\bar{p}$ production in $BB$ collisions
and the dashed arrow the threshold for $mB$ collisions, respectively.}
\label{bild3}
\end{figure}
\begin{figure}[h]
\centerline{\psfig{file=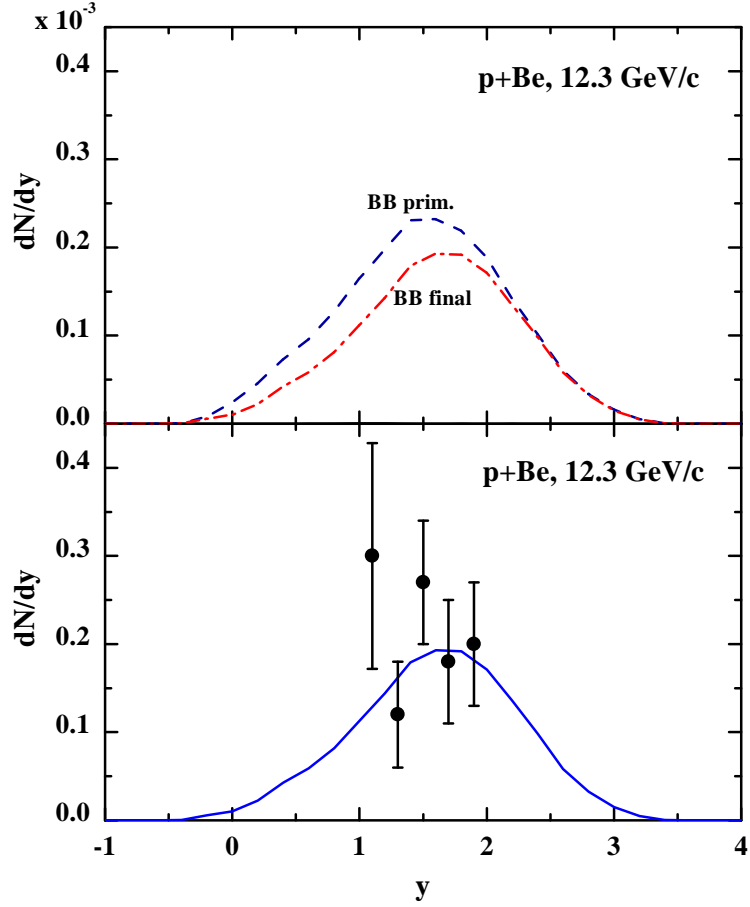,height=12cm}} \caption{Upper
part: Antiproton rapidity distribution in $p+Be$ reactions at 12.3
GeV/c within the HSD approach (dashed line, $BB prim$) without any
final state interactions. The influence of rescattering and
$\bar{p}$ annihilation is demonstrated in comparison to the
dash-dotted line ($BB final$), which shows the final antiproton
rapidity distribution. Note that there is no longer $\bar{p}$
annihilation for $y\geq$ 2 for the formation time $\tau_F$ = 0.8
fm/c adopted in the calculations. The contribution from $mB$
channels to antiproton production in this system is not visible on
a linear scale. Lower part: Comparison of the calculated $\bar{p}$
rapidity distribution (for $\tau_F$ = 0.8 fm/c) with the
experimental data from Ref. \cite{E910}. Within the statistics
achieved in the transport calculations  there is no change in the
$\bar{p}$ rapidity distribution when changing $\tau_F$ in the
interval $[0.4,1.6]$ fm/c.} \label{bild4}
\end{figure}
\begin{figure}[h]
\centerline{\psfig{file=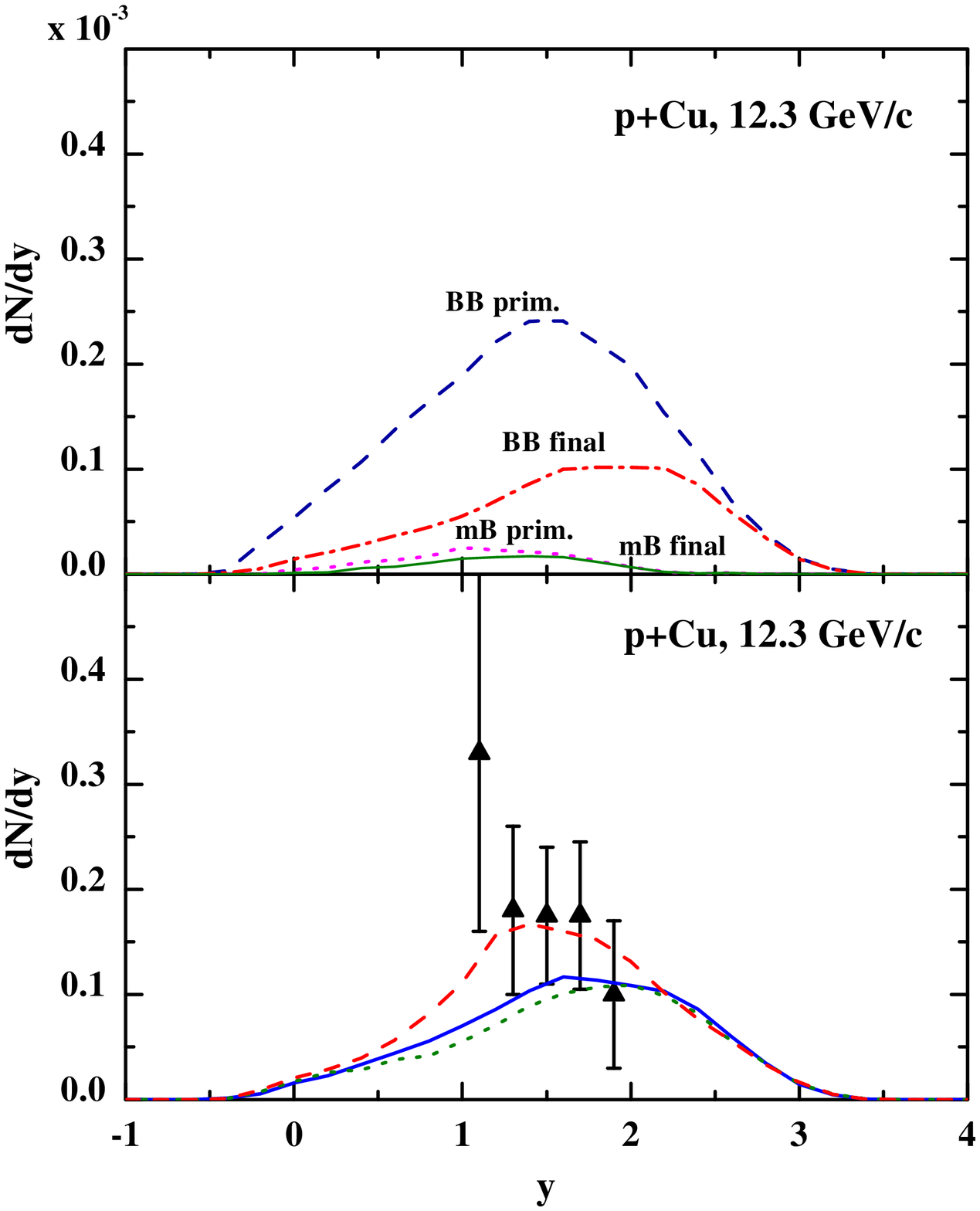,height=14cm}}
\caption{Upper part: Antiproton rapidity distribution in $p+Cu$ reactions at 12.3 GeV/c within the HSD
approach for baryon-baryon production channels (dashed line, $BB prim$) without any final
state interactions. The influence of rescattering and
$\bar{p}$ annihilation is demonstrated in comparison to the dash-dotted line ($BB final$), which
shows the final antiproton rapidity distribution from $BB$ channels for a formation time $\tau_F$ = 0.8 fm/c.
The short dashed line displays the contribution from secondary meson-baryon ($mB$)  reactions without FSI while
the thin solid line corresponds to the final $\bar{p}$ rapidity distribution from $mB$ reactions
including rescattering
and annihilation.
Lower part: Comparison of the calculated final $\bar{p}$ rapidity distribution from $BB$ and $mB$ collisions
(solid line for $\tau_F$ = 0.8 fm/c)
with the experimental data from Ref. \cite{E910}. The dashed and dotted lines demonstrate the results for
$\tau_F$ = 0.4 fm/c and 1.6 fm/c, respectively.}
\label{bild5}
\end{figure}
\begin{figure}[h]
\centerline{\psfig{file=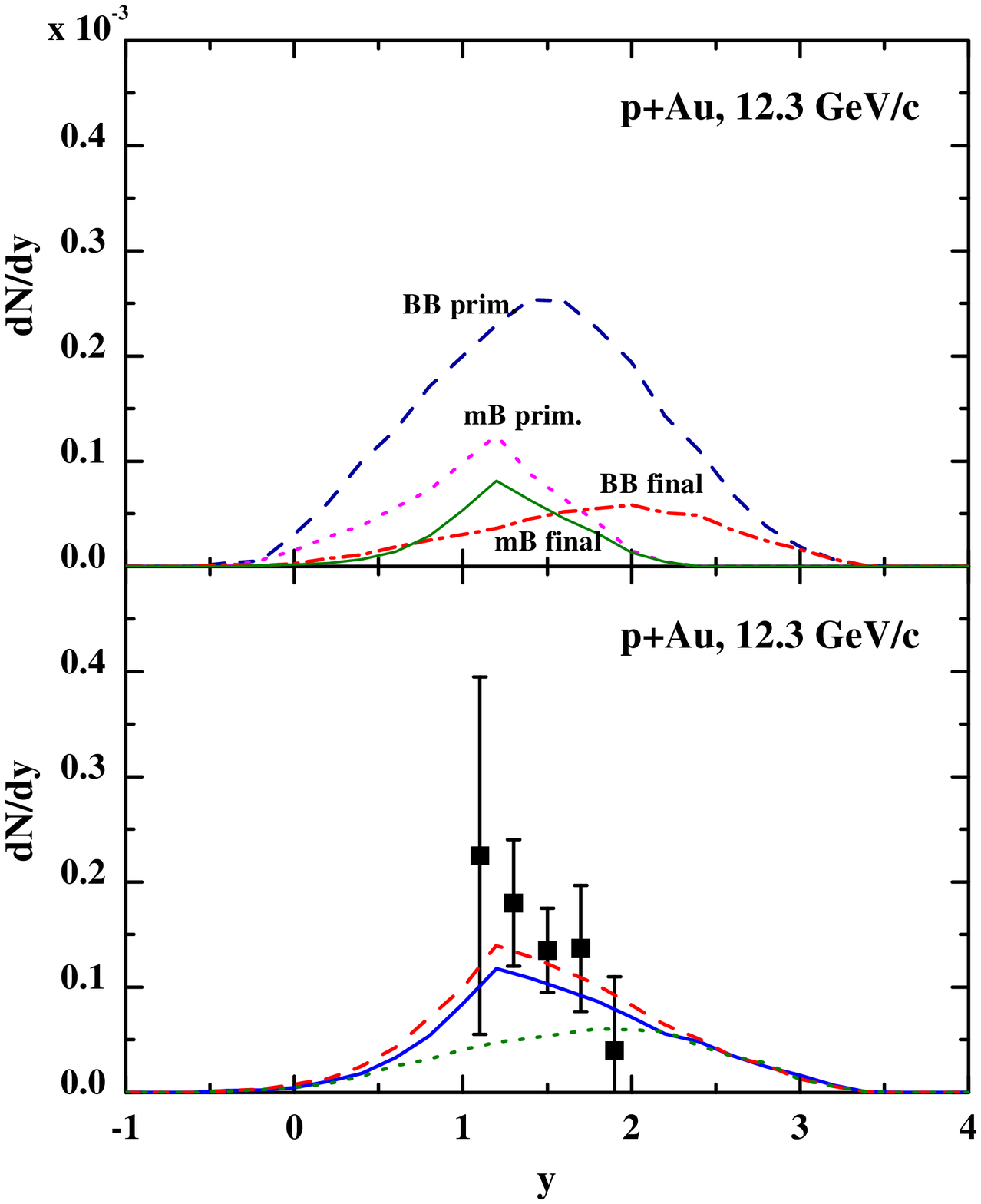,height=14cm}}
\caption{Upper part: Antiproton rapidity distribution in $p+Au$ reactions at 12.3 GeV/c within the HSD
approach for baryon-baryon production channels (dashed line, $BB prim$) without any final
state interactions. The influence of rescattering and
$\bar{p}$ annihilation is demonstrated in comparison to the dash-dotted line ($BB final$), which
shows the final antiproton rapidity distribution from $BB$ channels for a formation time $\tau_F$ = 0.8 fm/c.
The short dashed line displays the contribution from secondary meson-baryon ($mB$)  reactions without FSI while
the thin solid line corresponds to the final $\bar{p}$ rapidity distribution from $mB$ reactions
including rescattering
and annihilation.
Lower part: Comparison of the calculated final $\bar{p}$ rapidity distribution from $BB$ and $mB$ collisions
(solid line for $\tau_F$ = 0.8 fm/c)
with the experimental data from Ref. \cite{E910}. The dashed and dotted lines demonstrate the results for
$\tau_F$ = 0.4 fm/c and 1.6 fm/c, respectively.}
\label{bild6}
\end{figure}
\begin{figure}[h]
\centerline{\psfig{file=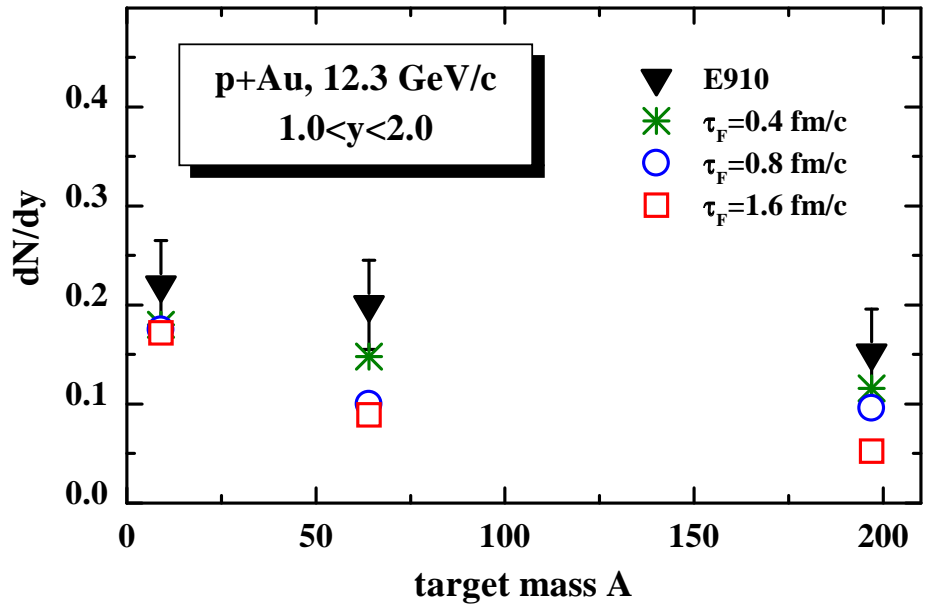,height=8cm}} \caption{
Antiproton rapidity distribution in $p+A$ reactions at 12.3 GeV/c
within the HSD approach for different formation times $\tau_F$ =
0.4, 0.8 and 1.6 fm/c for $Be$, $Cu$ and $Au$ targets. The
experimental data (full triangles) have been adopted from Ref.
\cite{E910}.}
 \label{bild7}
\end{figure}

\begin{figure}[h]
\centerline{\psfig{file=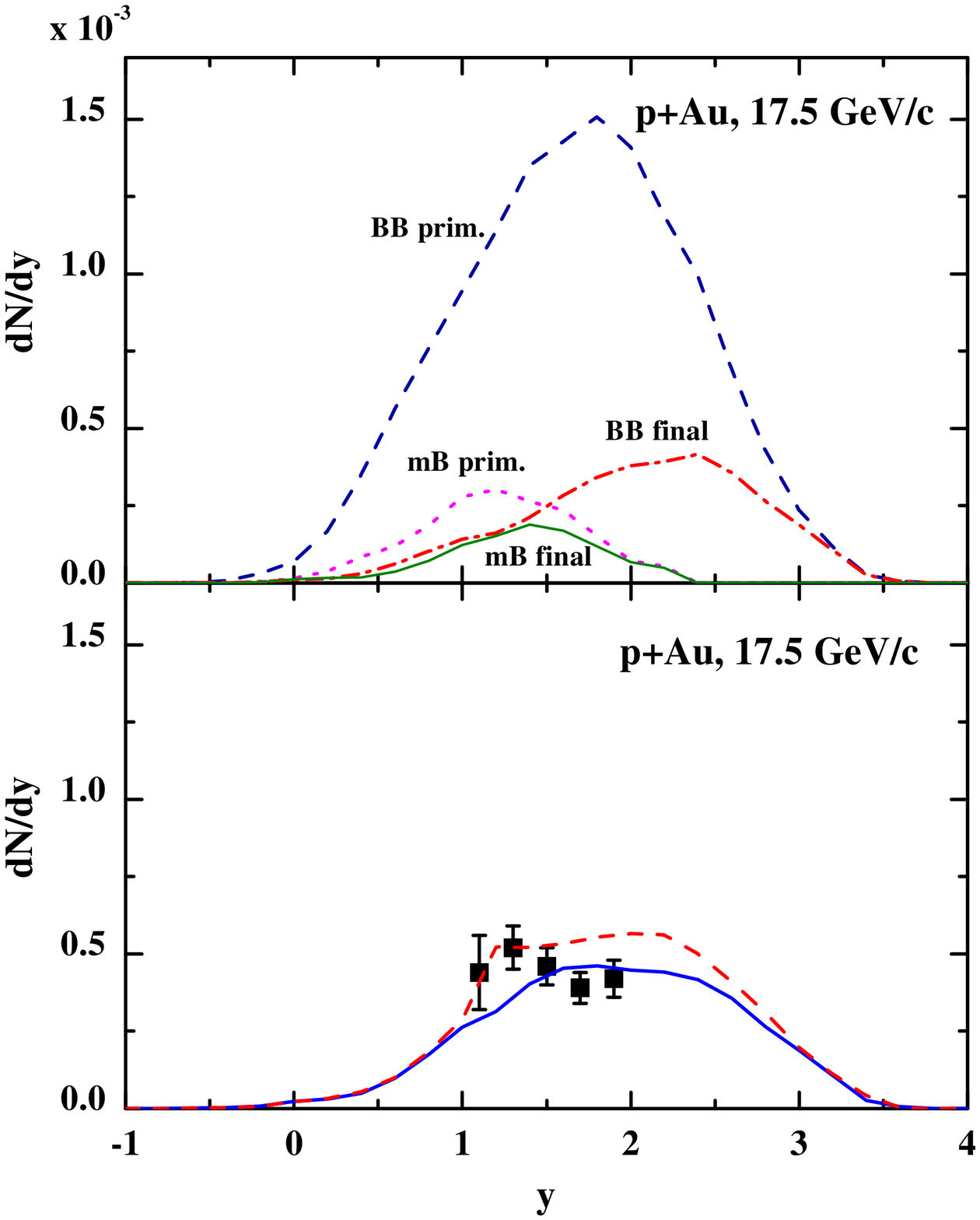,height=14cm}} \caption{Upper
part: Antiproton rapidity distribution in $p+Au$ reactions at 17.5
GeV/c within the HSD approach for baryon-baryon production
channels (dashed line, $BB prim$) without any final state
interactions. The influence of rescattering and $\bar{p}$
annihilation is demonstrated in comparison to the dash-dotted line
($BB final$), which shows the final antiproton rapidity
distribution from $BB$ channels for a formation time $\tau_F$ =
0.8 fm/c. The short dashed line displays the contribution from
secondary meson-baryon ($mB$)  reactions without FSI while the
thin solid line corresponds to the final $\bar{p}$ rapidity
distribution from $mB$ reactions including rescattering and
annihilation. Lower part: Comparison of the calculated final
$\bar{p}$ rapidity distribution from $BB$ and $mB$ collisions
(solid line: $\tau_F$ = 0.8 fm/c; dashed line: $\tau_F$ = 0.4
fm/c) with the experimental data from Ref. \cite{E910}.}
\label{bild8}
\end{figure}
\end{document}